\documentclass[prl,twocolumn,showpacs,superscriptaddress]{revtex4-1}
\usepackage{graphicx}
\usepackage{xcolor}
\usepackage{amsmath}
\usepackage{amssymb}
\usepackage{epsfig}
\usepackage{bm}
\usepackage[normalem]{ulem}
\usepackage[colorlinks,linkcolor=blue,anchorcolor=blue,citecolor=blue,urlcolor=blue]{hyperref}
\renewcommand{\v}[1]{{\bf #1}}
\newcommand{\Eq}[1]{Eq.~(\ref{#1})}
\newcommand{\Fig}[1]{Fig.\ref{#1}}
\newcommand{\Ref}[1]{Ref.\cite{#1}}
\newcommand{\<}{\langle}
\renewcommand{\>}{\rangle}

\newcommand{\cV}{{\cal V}}
\newcommand{\cE}{{\cal E}}

\newcommand{\ua}{\uparrow}
\newcommand{\da}{\downarrow}
\newcommand{\ra}{\rightarrow}

\newcommand{\twisting}{\eta}
\newcommand{\kv}{{\bf k}}

\newcommand{\sx}{\sigma_x}

\newcommand{\sz}{\sigma_z}

\newcommand{\sign}{{\rm sign}}
\newcommand{\w}{{\omega}}
\newcommand{\al}{\alpha}
\newcommand{\bt}{\beta}
\newcommand{\del}{\delta}
\newcommand{\Del}{\Delta}
\newcommand{\eps}{\epsilon}

\newcommand{\ga}{\gamma}
\newcommand{\Ga}{\Gamma}
\newcommand{\ka}{\kappa}

\newcommand{\La}{\Lambda}

\renewcommand{\th}{\theta}
\renewcommand{\dag}{\dagger}

\newcommand{\si}{\sigma}
\newcommand{\Si}{\Sigma}

\newcommand{\hybri}{{\mathcal{D}}(\omega) }
\newcommand{\hybr}{{\mathcal{D}}}
\newcommand{\SM}{Supplementary Materials}

\usepackage{color}

\newcommand{\eq}{\begin{equation}}
\newcommand{\ee}{\end{equation}}
\newcommand{\nn}{\nonumber\\}
\def\eqa{\begin{eqnarray}}
\def\eea{\end{eqnarray}}

\begin{document}

\title{Quantum impurities in channel mixing baths}

\author{Jin-Guo Liu}
\affiliation{National Laboratory of Solid State Microstructures $\&$ School of Physics, Nanjing
University, Nanjing, 210093, China}

\author{Da Wang}
\affiliation{National Laboratory of Solid State Microstructures $\&$ School of Physics, Nanjing
University, Nanjing, 210093, China}

\author{Qiang-Hua Wang}
\email{qhwang@nju.edu.cn}
\affiliation{National Laboratory of Solid State Microstructures $\&$ School of Physics, Nanjing
University, Nanjing, 210093, China}
\affiliation{Collaborative Innovation Center of Advanced Microstructures, Nanjing University, Nanjing 210093, China}


\begin{abstract}
    We propose a versatile strategy for numerical renormalization group solution of general channel-mixing Kondo and Anderson models beyond previous reach, opening the door toward broad applications in protocol non-perturbative machineries, such as dynamical cluster approximation and cluster dynamical mean field theory, for strongly correlated electron systems. We illustrate the strategy by investigating the quantum phase transitions in two quantum impurity models with cases untouched before.
\end{abstract}

%

\maketitle

The Kondo impurity model and its associated Anderson impurity model are challenging many-body problems and play an important role in condensed matter physics.\cite{Hewsonbook} Such models develop infrared logarithmic divergence in perturbation theory, and hence require non-perturbative solutions. The Kondo impurity model was first solved by Wilson using the simultaneously invented numerical renormalization group (NRG). \cite{wilson} The NRG captures the physics at exponentially decreasing energy scales iteratively, a key ingredient behind its great success. It proves to be one of the most accurate and efficient methods for the quantum impurity problems, \cite{nrg} and therefore has been broadly used as an impurity solver in the dynamical mean field theory (DMFT) and dynamical cluster approximation (DCA). \cite{dmftprl,dmft,dca,selfeng}

The quantum impurity is coupled to a noninteracting conduction band (or bath), the effect of which is to cause a so-called hybridization function $\hybri$. In the case of    a scalar or diagonal matrix function $\hybri$, a standard NRG procedure has been developed. \cite{mapping,oliveira,fixed,adaptive} The key ingredient is to map the effect of $\hybri$ into an open Wilson chain. However, the extension to quantum impurity models with $\hybri$ non-diagonal in the spin and/or local orbital basis is not straightforward at all. Such a matrix function is said to be channel-mixing henceforth, with a channel referring to a combination of spin and orbital (as well as atomic site in a cluster). This situation naturally arises in the presence of spin-orbital coupling, Cooper pairing, as well as in a cluster impurity. The intertwining spin, charge and orbital degrees of freedom are promising for harvesting novel quantum effects, and the cluster impurity is invoked in DCA and cluster DMFT (cDMFT). These interesting and important cases are barely addressed by NRG in the literature so far because of the lack of a versatile scheme to map a channel-mixing $\hybri$ to a Wilson chain. The only exception to our knowledge, in fact a special one of the general cases, is when $\hybri$ can be diagonalized by a frequency-independent unitary transformation in the presence of particular symmetries.\cite{sup1,*sup2,sup3,negativeUswave,KLMswave,Mitchell1,Mitchell2,Mitchell3,Mitchell4}  We also notice that the channel-mixing model we will be addressing is different to the usual multi-channel model where $\hybri$ is actually channel-diagonal, although the latter is interesting on its own right.\cite{multichannel}

In this Letter, we fill the gap caused by the above difficulties. We propose a versatile scheme to map a general (matrix) hybridization function into an open Wilson chain. We benchmark the mapping scheme against a nontrivial previous result,\cite{sup1,*sup2,sup3} and we illustrate the scheme for two channel-mixing cases untouched before. Our strategy enables NRG solution of general channel-mixing Anderson and Kondo models, as would be desirable in important protocols for strongly correlated electrons, such as DCA and cDMFT.

{\it A versatile mapping scheme}: For definiteness we consider a generalized Anderson impurity model, while the Kondo impurity model will be addressed in the closing section.
The Hamiltonian $H=H_{imp} + H_b + H_{hyb}$ is composed of,
\eqa && H_{imp} = \sum_{\al,\bt} f^\dag_\al h_{loc}^{\al\bt}f_\bt + H_{int},\nn
     && H_b = \sum_{\v k,a,b} c^\dag_{\v k, a} h_\v k^{ab} c_{\v k, b},\\
     && H_{hyb} = \frac{1}{\sqrt{N}}\sum_{\v k,\al,a} (f^\dag_\al \ga^{\al a}_\v k c_{\v k,a} + {\rm h.c.}).\nonumber
\eea
Here $f_\al$ is an annihilation field at channel $\al$ of the impurity, $c_{\v k,a}$ an annihilation field at channel $a$ and at momentum $\v k$ in the bath, and $N$ is the volume (or number of unit cells) in the bath. (The channel numbers in the impurity and the bath can differ in general.) As we mentioned, we take a channel index as a combined label of spin and orbital (as well as the site in a cluster impurity). The concrete expressions for the matrices $h_{loc}$, $h_\v k$ and $\ga_\v k$, as well as the interaction $H_{int}$ on the impurity are unnecessary at this stage.

Integrating out the free bath leads to a (matrix) self-energy correction to the impurity,
\eqa \Si(z) = \frac{1}{N}\sum_\v k \ga_\v k G_{\v k}(z) \ga^\dag_\v k,\eea
where $z=\w\pm i0^+$ in the retarded/advanced case, and $G_{\v k}(z)=1/(z - h_\v k)$ is the Green's function at the complex frequency $z$. Via Kramers-Kronig relation, $\Si(z)$ can be completely characterized by
\eqa \hybri = i [\Si(\w+i0^+)-\Si(\w-i0^+) ]/2\pi.\label{defhyb} \eea
which we call a (matrix) hybridization function. (For later convenience our definition differs to the usual one by a factor of $\pi$.) However, $\hybri$ is in general non-diagonal, i.e., channel mixing. Our purpose is to map the effect of such a $\hybri$ into that of a generalized open Wilson chain.

By definition, $\hybri$ is hermitian and positive semi-definite. This enables us to write
\eqa \hybri = \sum_n |n,\w\>\rho_n(\w)\<n,\w|,\eea
where $\rho_n(\w)\geq 0$ and $|n,\w\>$ are the $n$-th eigenvalue and eigenvector of $\hybri$. Similarly to the usual one-band case,\cite{oliveira,fixed} we can re-express $\rho_n(\w)$ as \eqa \rho_n(\w)=\int dx~ t_n^2 (x) \del[\w-\eps_n(x)],\eea
where $\eps_n(x)$ is a continuous function of $x$ for each $n$, subject to the requirement on the integration measures
\eqa  t_n^2(x) ~|dx| = \rho_n[\eps_n(x)] ~|d\eps_n(x)|. \label{jacobi} \eea
[The modulus symbol is necessary if $d\eps_n(x)/dx<0$.]
The parametrization of $\eps_n(x)$ and $t_n(x)$ for $\rho_n(\w)$ is similar to the usual logarithmic way,\cite{oliveira,fixed,nrg} and is also provided in the \SM.

\begin{figure}
\includegraphics[width=7.8cm]{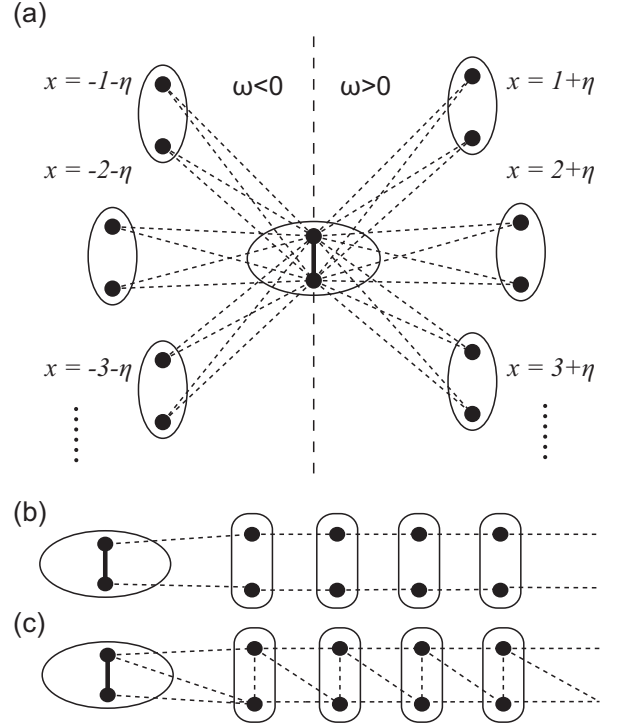}
\caption{ Graphical illustrations of (a) the discretized bath coupled to the impurity, (b) a channel-diagonal Wilson chain in the usual case, and (c) a channel-mixing Wilson chain in general cases. The number of channels is set as (but not limited to) $I=2$.} \label{illustrate}
\end{figure}

Substituting the expression of $\rho_n(\w)$ into $\hybri$, we write
\eqa \hybri && =  \int dx ~ \sum_n |n,\w\> t_n^2(x)\del[\w-\eps_n(x)]\<n,\w| \nn
            && =  \int dx ~ \sum_n |n,x\> t_n^2(x)\del[\w-\eps_n(x)]\<n,x|, \nonumber
\eea
where we made a replacement $|n,\w\> \ra |n,x\>\equiv |n,\eps_n(x)\>$, valid in the presence of the delta-function in the integrand. This is a crucial step for the following discussions. In components, we have \eqa {\hybr}_{\al\bt}(\w) =  \int dx ~ \sum_n V_{\al n}(x) \del[\w-\eps_n(x)] V^\dag_{n\bt}(x), \nonumber\eea
where $V_{\al n}(x)=t_n(x)\<\al|n,x\>$. The above parametrization maps $H_b$ and $H_{hyb}$ to, respectively,
\eqa && H'_b = \int dx \sum_n \psi_n^\dag (x) \eps_n(x) \psi_n(x), \nn
     && H'_{hyb} = \int dx \sum_{\al,n} f^\dag_\al V_{\al n}(x) \psi_n(x) +{\rm h.c.},\nonumber
\eea
where $\psi$'s are fermion fields in $x$. The mapping is exact since integrating out $\psi$'s leads to the same $\Si(z)$ we started from.
In practice, however, one would like to proceed in a discrete space of $x$. Fortunately, as in the usual case, the above mapping is designed such that its discretized version provides a good approximation. For example, one can take $x_j = (|j|+\twisting)\sign(j)$\cite{oliveira} with nonzero integer $j$ and a real twisting factor $\twisting\in (0,1]$, and replace the integration over $x$ to a summation over $j$. Each case of $\twisting$ represents an approximation of the continuum model, and as far as $\hybri$ is concerned, the average over $\twisting$ reproduces exactly the result from the continuum model. A graphical representation of a discrete mapping is shown in \Fig{illustrate}(a).

We proceed to map the discretized model to an open Wilson chain, a key ingredient of NRG to reduce the computational cost and optimize the scaling behavior. 
In the special case when $|n,x\>=|n\>$ is independent of $x$, a unitary transformation $\sum_\al f_\al^\dag\<\al|n\>\ra f_n^\dag$ makes $H'_b+H'_{hyb}$ diagonal in $n$ (while $H_{imp}$ may not and actually does not have to be so), a case accessible by the conventional mapping scheme.\cite{sup1,*sup2,sup3,negativeUswave,KLMswave,Mitchell1,Mitchell2,Mitchell3,Mitchell4} But this does not apply to more general occasions we are aiming at. Let us assume the dimension of the single-particle Hilbert space of the impurity is $I$ (the number of channels), and that for $H'_b$ is $B=2JI$, where $J$ is the number of $x_j$'s for $j>0$ (or $j<0$) retained in the discritization. Let us rewrite $H'_b$ and $H'_{hyb}$ compactly as
\eqa H'_b = \Psi^\dag \cE \Psi,\ \ \ H'_{hyb} = F^\dag \cV \Psi + \Psi^\dag \cV^\dag F,\nonumber \eea
where $\Psi$ and $F$ are spinors composed of the $\psi$- and $f$-fields, respectively.
In this form, $\cE_{B\times B}$ and $\cV_{I\times B}$ are matrices. We perform a QR decomposition $\cV^\dag = U_0^\dag T_0^\dag$, where $T_0^\dag$ is an upper triangular matrix. We use the columns of $U_0^\dag$ as the first set of Krylov basis vectors to transform $\cE$, by block-Lanczos,\cite{wiki,blocklanczos} into a block-tridiagonal matrix $E_T$ (with block size $I$) such that $\cE=U^\dag E_T U$, where the columns of $U^\dag$ are Krylov basis vectors, and the leading ones are from $U_0^\dag$. In the resulting Krylov space, or upon a canonical transformation $U\Psi\ra\Phi$, we have
\eqa  H'_b+H'_{hyb}\ra \sum_{k=1}^{K}[\Phi_k^\dag E_k \Phi_k+\Phi^\dag_{k-1}T_{k-1}\Phi_{k}+{\rm h.c.}], \label{chain}
\eea
where $K$ is the number of block-Lanczos iterations, $\Phi_k$ denotes an $I$-component spinor such that $\Phi^\dag=(\Phi^\dag_1,\Phi_2^\dag,\cdots,\Phi^\dag_{K})$, and we set $\Phi_0 = F$ for brevity. For $k\geq 1$, $E_k$ ($T_k$) is the $k$-th $I\times I$ block element of $E_T$ along the diagonal (upper sub-diagonal). Notice that all $T_k$'s are triangular matrices themselves. \Eq{chain} defines the open Wilson chain we were after. For comparison, the usual Wilson chain is illustrated in \Fig{illustrate}(b), which is channel diagonal, while the Wilson chain in our general cases is illustrated in (c). In practice, the block-Lanczos procedure requires infinitely high precision and is truncated at a suitable stage $K \sim J$. Numerical examples of our mapping scheme for models discussed below can be found in the \SM.

{\it Quantum impurity in an $s$-wave superconducting bath}:  Given the above versatile mapping, we are able to study any quantum impurity models, with or without channel mixing. Here we consider an Anderson impurity coupled to a conventional $s$-wave superconductor. The Hamiltonian in the Nambu space is composed of
\eqa && H_{imp}=\psi_f^\dag\epsilon_f\sigma_z\psi_f -\frac{1}{2}U(\psi_f^\dag\si_z\psi_f)^2,\nn
     && H_{b}=\sum_\kv \psi_{\kv}^\dag (\epsilon_{\kv}\sz+\Delta_{\kv}\sx)\psi_{\kv},\nn
     && H_{hyb}=\frac{1}{\sqrt{N}}\sum_\kv \psi_f^\dag t_{\kv}\sz \psi_{\kv}+{\rm h.c.}.\nonumber
\eea
Here $\psi_f^\dag=(f_\ua^\dag,f_\da)$ and $\psi^\dag_\kv =(c_{\kv\ua}^\dag,c_{-\kv\da})$ are Nambu spinors for the impurity and the bath, $\si_{x,z}$ are Pauli matrices in the Nambu space, $\eps_f$ is a measure of the deviation from particle-hole symmetry on the impurity, $U$ is the Hubbard repulsion, $\Del_\kv$ is the pairing function of momentum $\kv$ in the bath, and finally $t_\v k$ is the momentum-dependent coupling amplitude between the impurity and the bath. We assume $\Del_\kv=\Del$ and $t_\kv=t$ as in \Ref{sup1,*sup2,sup3}.  However, we assume a more general normal state density of states (DOS) $\rho(\w)=\rho_0(1+\ka\w)$ for $\w\in [-D_0,D_0]$ and $\rho_0=1/2D_0$. Notice that $\ka$ is a measure of the asymmetry in the DOS. Henceforth we take $D_0=1$ as the unit of energy. Integrating out the superconducting bath, we get $\Si(z)$ and subsequently
\eqa \hybri =\frac{\Ga [|\w|\si_0+\ka\w \xi\si_z-\Del\si_x\sign(\w)]}{\pi \xi}W(\xi^2),\nonumber \eea
where $\Ga=\pi\rho_0 t^2$, $\si_0$ is the $2\times 2$ identity matrix, and we use $\xi=\sqrt{\w^2-\Del^2}$ for brevity. Henceforth we use $W(u)=\th(u)\th(1-u)$ as a window function.

\begin{figure}
\includegraphics[width=8.5cm]{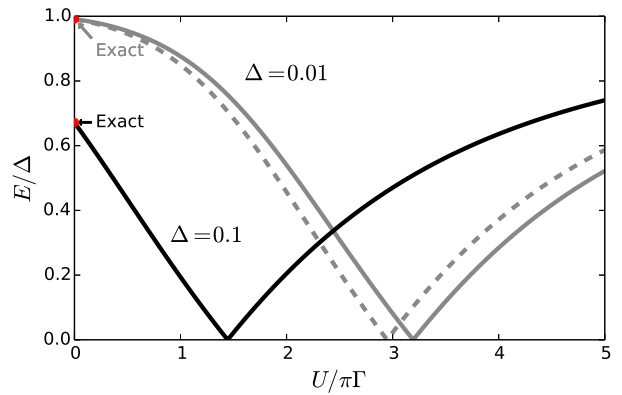}
\caption{The first excitation energy $E$ as a function of $U$ for $\Delta=0.01$ (gray solid line) and $\Delta=0.1$ (black solid line). The dashed line is the result for $\Del=0.01$ but with proper accounting of the under-estimation in \Ref{sup3} for comparison. See the text for details.} \label{abs}
\end{figure}

First we consider $\ka=0$ and $\eps_f=0$, a case studied in \Ref{sup1,*sup2,sup3}, but we proceed using the present scheme for comparison. We map the system to a Wilson chain as described above, and we perform NRG iterations thereafter.We monitor the energy $E$ of the lowest excited state on top of the many-body ground state as a function of $U$. The results are shown in \Fig{abs} for $\Del=0.01$ (gray solid line) and $\Del=0.1$ (black solid line). The NRG results in the free limit $U=0$ is in excellent agreement with the analytical results (arrows) from the Green's function method. With increasing $U$, we see, in either case of $\Del$, a cusp at which $E=0$. It is known that this corresponds to a singlet-doublet transition of the many-body ground state. Our transition point for $\Del=0.01$ (gray solid line) is slightly larger than that in \Ref{sup3}. This is however not an inconsistency. In fact, the discretization scheme in \Ref{sup3} under-estimates the effect of $\Ga$ by a factor of $A=\frac{1}{2}\frac{\La+1}{\La-1}\ln\La$, \cite{oliveira,nrg} where $\La$ is a scaling factor. For  $\La=2.5$ used in \Ref{sup3}, $A\sim 1.069$. Thus for a fair comparison, we use $\Ga'=\Ga/A$ in our calculation, and present the data at $\Ga$ instead. The result is shown as the dashed line in \Fig{abs}, which is now in agreement with that in \Ref{sup3}.

\begin{figure}
\includegraphics[width=8.5cm]{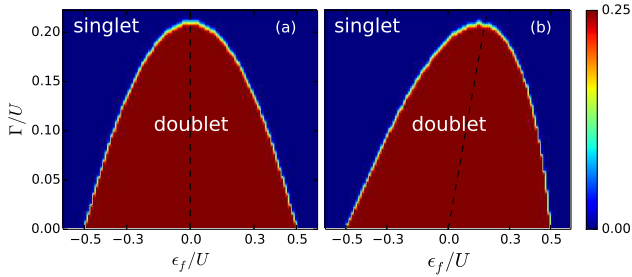}
\caption{The phase diagram for an $s$-wave superconducting bath. (a) $\ka=0$ and (b) $\ka=1$. In both cases $\pi\Ga=0.5$ and $\Del=0.1$. The color indicates the magnitude of $M^2$. At zero temperature, the phase boundary would be sharp and mark the transition from doublet to singlet ground states. }\label{phase_s}
\end{figure}

Next we consider both $\ka=0$ and $\ka=1$, and ask how $\ka$ influences the ground state. {\it The case of $\ka\neq 0$ is beyond previous reach, but no excess difficulty arises in our strategy.} For our purpose we calculate the static magnetic susceptibility $\chi_{\rm imp}(T)=\chi_{\rm tot}(T)-\chi_{\rm tot}^{(0)}(T)$ at an energy scale $T\sim 10^{-6}$ (the zero temperature limit).\cite{nrg} Here, $\chi_{\rm tot}$ ($\chi_{\rm tot}^{(0)}$) is the total susceptibility calculated with (without) the impurity. An effective moment scale $M$ can be defined by $M^2 = T\chi_{imp}(T)$. One expects $M^2$ to vanish if the impurity spin is screened, while $M^2=1/4$ if a full local moment survives. In fact, they correspond to the singlet and doublet ground states, respectively. The resulting phase diagram revealed by $M^2$ is shown in \Fig{phase_s}. While the phase boundary is symmetric in $\eps_f$ for $\ka=0$ in (a), it is boosted toward the right for $\ka=1$ in (b). This can be understood as follows. Given the above $\hybri$, it can be shown that $\ka$ {\it alone} leads to a self-energy correction $\del\eps\si_z$ as $\w\ra 0$, with $\del\eps=-2\ka\Ga(\sqrt{1+\Del^2}-|\Del|)/\pi$. This means the effective impurity level is shifted roughly by $\del\eps$. Therefore, if $\Ga/U=g(\eps_f/U)$ describes the phase boundary for $\ka=0$, it becomes $\Ga/U=g[(\eps_f+\del\eps)/U]$ for $\ka\neq 0$. This explains qualitatively the behavior of the phase boundary in \Fig{phase_s}(b).

{\it Quantum impurity in a $d$-wave superconducting bath}: We now consider a $d$-wave superconducting bath. For simplicity, we use the same normal band with $\ka=0$ as before. In case of constant $t_\kv$, $\hybri$ is actually diagonal, the effect of which is equivalent to that of a nodal normal metal that can not screen the impurity spin at $\eps_f=0$.\cite{sup3} In this respect, it is interesting to ask how an off-diagonal (channel-mixing) term in $\hybri$ would occur and how it would affect the fate of impurity spin at zero temperature. {\it This is an issue not yet addressed.}

In fact, to induce channel-mixing in our case, all that we need is a $t_\kv$ asymmetric under the point group. For example, in a lattice model of the bath, if the impurity is coupled, via hopping integral $t$, only to two sites at $\v r=\pm \hat{x}/2$ on a nearest-neighbor bond, we would have $t_\v k=2t\cos(k_x/2)$ (in a suitable gauge). This can be resolved into $A_{1g}$ and $B_{1g}$ lattice harmonics. Thus in general, we may assume $t_\v k\ra t_\phi = \sum_l t_l e^{il\phi}$ in the continuum limit, with $\phi$ the azimuthal angle of $\kv$, and $t_{-l}=t_l^*$ required by time-reversal symmetry. In the same limit we write $\Del_\kv \ra \Del_\phi =\Del\cos(2\phi)$. For concreteness, we assume $t_\phi=t_0+2t_2\cos(2\phi)$, and we fix $\Del=0.1$ in the calculations.

\begin{figure}
\includegraphics[width=8.5cm]{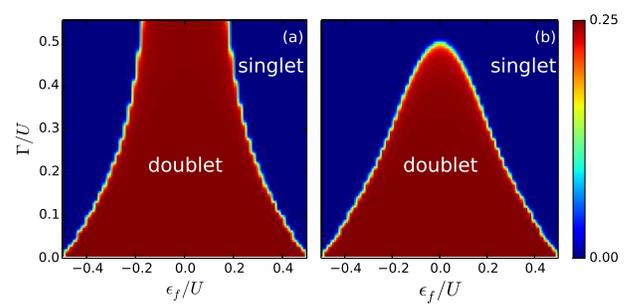}
\caption{The phase diagram for a $d$-wave superconducting bath. (a) $t_2/t_0=0$ and (b) $t_2/t_0=0.1$. In both cases $t_0$ is fixed by $\pi\Ga=0.5$. The color indicates the magnitude of $M^2$.  }\label{phase_d}
\end{figure}

The $d$-wave superconducting bath leads to
\eqa \hybri =\int\frac{d\phi}{2\pi}\frac{\Ga |t_\phi|^2 (\w\si_0-\Delta_\phi\sigma_x)\sign(\w)}{\pi |t_0|^2\xi_\phi}W(\xi_\phi^2),\nonumber
\eea
where $\Ga=\pi\rho_0 t_0^2$ and $\xi_\phi=\sqrt{\w^2-\Del_\phi^2}$. We make the resolution $\hybri=d_0(\w)\si_0+d_x(\w)\si_x$ to find, for $|\w|\ll 1$, $d_0(\w)\sim (t_0^2+2t_2^2)|\w|/\Del$ and $d_x(\w)\sim t_0 t_2 \w|\w|/\Del^2$. Thus $t_2$ leads to a channel-mixing component $d_x(\w)$. We perform NRG using the present strategy. \Fig{phase_d} shows the resulting phase diagram revealed by $M^2$, with $\pi\Ga=0.5$ fixed. There is a marked difference when $\eps_f=0$, the particle-hole symmetric point for the impurity: while a full local moment persists for $t_2/t_0=0$ in (a), we do find a transition from the local moment phase to complete screening as $\Ga/U$ increases for $t_2/t_0=0.1$ in (b), similarly to the case of $s$-wave bath in \Fig{phase_s}(a). Indeed, the behavior of $d_x(\w)$ is such that it vanishes for small $|\w|$ but is finite above the gap edge, resembling qualitatively that from an $s$-wave bath. This indicates the essential role played by an asymmetric $t_\kv$.

{\it Summary and remarks}: We developed a versatile scheme to map a general quantum impurity model to an open Wilson chain, resolving the difficulty caused by channel mixing. This opens the door toward broad applications in protocol nonperturbative machineries, such as DCA and cDMFT, for strongly correlated electron systems. We illustrated the strategy by investigating the quantum phase transitions in two quantum impurity models with cases untouched before.

We remark how our mapping scheme also applies to general Kondo impurity models. Suppose the bath is composed of two parts $A$ and $B$, and only $A$ is coupled via spin-exchange to the quantum impurity $I$. The idea is to map the effect of $B$ on $A$, using our scheme, as an open Wilson chain $C$ that starts from $A$. Then $I\otimes C$ forms the suitable system for NRG iterations.

\acknowledgments{The project was supported by NSFC (under grant Nos.11574134 and 11504164) and the Ministry of Science and Technology of China (under grant No.2011CBA00108 and 2011CB922101). LJG thanks Rok $\check{\rm Z}$itko and Z. L. Gu for helpful discussions.}

\bibliography{ref}

\end{document}